\documentclass[a4paper]{article}

\usepackage{itgspeech2016}    
\usepackage{times}            
\usepackage[english]{babel}   
\usepackage[ansinew]{inputenc}
\usepackage[T1]{fontenc}      
\usepackage[sort&compress,numbers]{natbib}	
\usepackage{amsmath,amssymb}
\usepackage{graphicx}
\usepackage[colorlinks=false,pdfborder={0 0 0}]{hyperref}
\usepackage{units}
\usepackage{psfrag}
\usepackage[mathscr]{euscript}
\usepackage{color}

\title{Unsupervised Classification of Voiced Speech and Pitch Tracking Using \\Forward-Backward Kalman Filtering}

\author{Benedikt Boenninghoff$^{~1}$, Robert M. Nickel$^{~2}$, Steffen Zeiler$^{~1}$, Dorothea Kolossa$^{~1}$}

\address{$^{1}$Institute of Communication Acoustics, Ruhr-Universit\"at Bochum, 44780 Bochum, Germany\\
  Email: \texttt{\{benedikt.boenninghoff,dorothea.kolossa,steffen.zeiler\}@rub.de}\\
  $^{2}$Department of Electrical and Computer Engineering, Bucknell University, Lewisburg, PA 17837, USA\\
  Email: \texttt{robert.nickel@bucknell.edu}\\
 }

\begin{document}

\maketitle

\begin{abstract}
The detection of voiced speech, the estimation of the fundamental frequency and the tracking of pitch values over time are crucial subtasks for 
a variety of speech processing techniques. Many different algorithms have been developed for each of the three subtasks. 
We present a new algorithm that integrates the three subtasks into a single procedure. The algorithm can be applied to pre-recorded 
speech utterances in the presence of considerable amounts of background noise. We combine a collection of standard metrics, 
such as the zero-crossing rate for example, to formulate an unsupervised voicing classifier. The estimation of pitch values 
is accomplished with a hybrid autocorre- lation-based technique. We propose a forward-backward Kalman filter 
to smooth the estimated pitch contour. In experiments we are able to show that the proposed method compares favorably with current, 
state-of-the-art pitch detection algorithms.
\end{abstract}

\vspace{-0.05in}

\section{Introduction}

Human speech can roughly be divided into voiced and unvoiced sounds. Unvoiced sounds are produced via a non-periodic/turbulent airflow along the vocal tract and the lips/ teeth. Voiced speech sounds are produced by the vibration of the vocal chords, where the airflow is interrupted periodically. The inverse of the duration between two interruption epochs is called the fundamental frequency or $F_0$~\cite{VaryMartin}. 

An accurate tracking of $F_0$ over time is still considered a major technical challenge. In general, speech is not perfectly periodic and the correct detection of voiced speech sounds as well as the 
accurate estimation of the pitch value is difficult, particularly when speech is recorded in a noisy environment.

When we estimate the pitch we usually do not only have a single voiced speech frame to work with but an entire spoken utterance. 
The task, then, is to track the pitch value throughout all voiced segments of the utterance. This can be divided into three subtasks: (a) detection of voiced sounds, (b) pitch estimation and (c) pitch tracking. 

(a) Different methods have been developed to 
distinguish between voiced and unvoiced or silent speech segments~\cite{V-UV,Kondoz}.
The decision can be made before the pitch estimation algorithm is applied or it can be coupled with the result of the pitch estimation~\cite{study1,study2,study3}.

(b) Various parametric as well as non-parametric pitch estimation methods have been studied, using properties of both time and frequency 
domain characteristics~\cite{study3}. 
The algorithms differ in terms of their signal model and the application of statistical tools to detect repeating patterns within a signal.
The autocorrelation function is a wide- spread tool to detect pitch within a signal in the time domain. Autocorrelation functions are applied in 
Praat~\cite{Praat} and Yin~\cite{YIN}. In the frequency domain, the cepstrum meth- od~\cite{RabinerSchafer}  and the harmonic periodic 
spectrum~\cite{HPS} can be used to estimate maximum pitch values. Hybrid methods such as Pefac~\cite{Pefac} and BaNa~\cite{BaNa} use properties 
in both domains. In~\cite{Theory} multi-pitch estimation algorithms embedded in a statistical framework are developed based on a harmonic 
constraint. It is assumed that the signal is composed of complex sinusoids. In~\cite{harmonic} a maximum likelihood (ML) pitch estimator is 
developed using a real-valued sinusoid-based signal model. Halcyon~\cite{Halcyon} presents an algorithm for instantaneous pitch estimation 
that decomposes the signal into narrow-band components and calculates fine pitch values based on the estimated instantaneous parameters.

(c) There exist several methods to find the correct pitch track for a speech sentence.
BaNa uses the Viterbi algorithm to find the best path of the given pitch candidates for each frame with respect to a cost function. 
In~\cite{harmonic} two approaches based on hidden Markov models and Kalman filters are proposed to track the ML pitch estimate.
A neural network based pitch tracking method is presented in~\cite{Deep}, where Pefac is used to generate feature vectors and again the 
Viterbi algorithm is applied to detect the optimal pitch sequence.

Even neglecting voiced speech classification and/or the training phase, some of the existing algorithms need long computation times.  The goal of this paper is to combine the three subtasks from above to formulate a computationally efficient voiced speech classifier and pitch tracking algorithm. The algorithm works reliably on speech utterances that have been pre-recorded in a noisy environment. A very important aspect is that \underline{no} training phase is needed. 
To that end, we have devised a method that autonomously detects voiced frames without a training phase. The problem is solved by combining different approaches from \cite{V-UV,Kondoz} to create a feature vector for unsupervised k-means clustering and subsequent linear discriminant analysis. 

Using the resulting voiced speech segments, we can estimate the corresponding pitch values. Here, we use a hybrid method based on the spectro-temporal autocorrelation function (SpTe-ACF) as proposed in \cite{Kondoz}. All estimated pitch values are then input to the proposed forward-backward Kalman filter (FB-KF) to obtain a smoothed pitch track for the entire speech utterance. 

\vspace{-0.05in}

\section{Classification of Voiced Speech}
\label{sec:classification}
Voiced and unvoiced speech segments have different characteristics regarding their periodicity, frequency content, spectral tilt, etc. which 
can be used for their distinction.

\subsection{Feature Extraction}
Let $x_k(n)$ with $0 \leq n \leq N-1$ be the $k$-th frame of length $N$. 
We combine several voiced speech classification methods to construct a feature vector $\boldsymbol{v}(k) = [v_1(k), \ldots ,v_D(k) ]^T$,
where each entry is the normalized result of one classification method.
Since some methods do not work if there is neither noise nor speech present in a frame, only speech frames with enough power are considered.
To detect silent frames, we compute $P_x(k) = \frac{1}{N}\sum_{n=0}^{N-1} x_k^2(n)$. 
For $P_x(k) < \delta_{P_x}$, our entire feature vector is set to $\boldsymbol{v}(k) = [-1, \ldots, -1 ]^T$.
All computed classification methods are 
normalized in such a way that they are restricted within the interval $[-1,1]$, where a value close to one is assigned as a voiced frame~\cite{Kondoz}.

In the following, we combine five different methods, namely the periodic similarity measure, zero-crossing rate, spectrum tilt, 
pre-emphasized energy ratio and low-band to full-band ratio as described in~\cite{Kondoz}.

\subsection{Unsupervised Decision Making}
The next task is to find a linear weighting vector $\boldsymbol{w}^T \boldsymbol{v}(k) = \sum_{d=1}^{D} w_d v_d(k)$ 
such that we can assign $\boldsymbol{v}(k)$ to one class $C_i$ with $i = \left\{ 0,1\right\}$, where $C_0$ denotes the set of unvoiced frames and 
$C_1$ is the set of voiced frames.
The class labels for all feature vectors can be obtained using k-means clustering. The 
algorithm is initialized with means $\boldsymbol{\mu}_0 = [-1 \ldots -1]^T$ and $\boldsymbol{\mu}_1 = [1 \ldots 1 ]^T$
so that the class allocation is unique. 
Having the class labels, we can apply a linear discriminant analysis to obtain the optimal weighting vector. 
For two classes it is simplified to~\cite{Theodoridis}
\vspace{-0.015in}
\begin{align}
    \label{eq:LDA}
    \boldsymbol{w} \propto \boldsymbol{S}_w^{-1} \big[ \boldsymbol{m}_0 - \boldsymbol{m}_1 \big],
\end{align} \vspace{-0.15in}

\noindent where the local mean vectors $\boldsymbol{m}_i$ with $i = \left\{ 0,1\right\}$ result from the last 
round of k-means clustering and $\boldsymbol{S}_w^{-1}$ denotes the inverse of the within-class scatter matrix.
As the $D$-dimensional scatter matrix is symmetric, there are at most $D (D+1) /2$ entries to compute. 
In our simulations, utterances with a length of at least $1$ second are sufficient to be able to compute the inverse of the 
within-class scatter matrix, provided that each class is sufficiently represented.
Now we can use the linear classifier of Eq.~(\ref{eq:LDA}) for the decision making,
\vspace{-0.2in}
\begin{align}
    \label{eq:decision}
    \boldsymbol{w}^T \boldsymbol{v}(k) \underset{\mathrm{unvoiced}}{\overset{\mathrm{voiced}}{\gtrless}} 0.
\end{align} \vspace{-0.1in}

\noindent Since class allocation is unique, the decision boundaries can be learned ad-hoc for each sentence separately, which provides two advantages: 
First, no training phase is needed. Second, the instantaneous adjustment of the classifier considers the current speaker as well as 
the noise type.

\begin{figure}[t]
    \psfrag{s01}[t][c][1.3]{time (s)}
    \psfrag{x01}[c][c][1.3]{$2.2$}
    \psfrag{x02}[c][c][1.3]{$2.5$}
    \psfrag{x03}[c][c][1.3]{$2.8$}
    \psfrag{x04}[c][c][1.3]{$3.1$}
    \psfrag{x05}[c][c][1.3]{$3.4$}
    \psfrag{x06}[c][c][1.3]{$3.7$}
    \psfrag{v01}[r][r][1.3]{-$1$}
    \psfrag{v02}[r][r][1.3]{$0~$}
    \psfrag{v03}[r][r][1.3]{$1~$}
 \begin{minipage}[b]{1\linewidth}
    \centering
    \centering{\scalebox{0.5}{\includegraphics{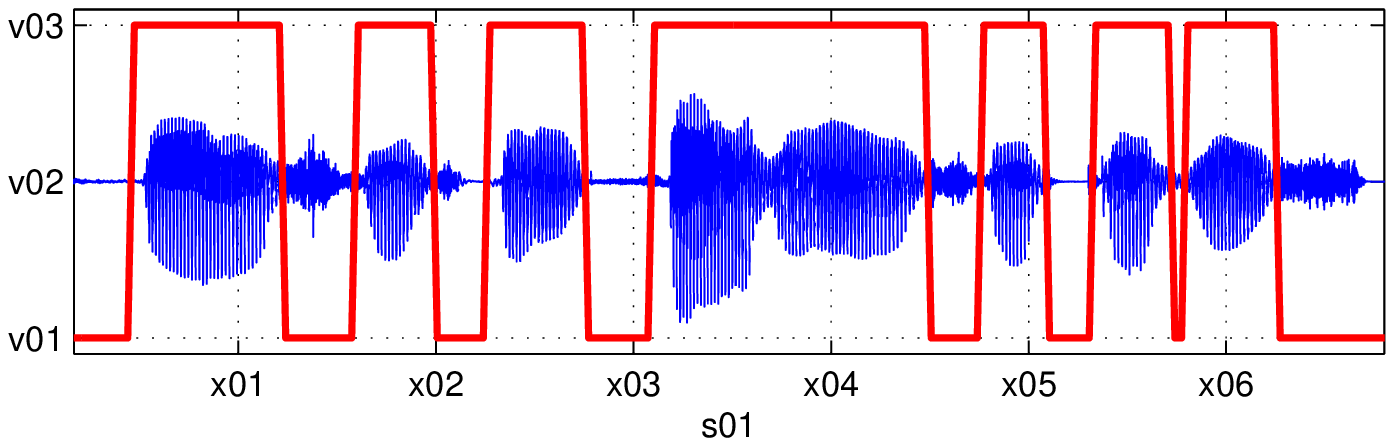}}}
    \centerline{(a) Noise-free speech sentence.}
\end{minipage}
\hfill
\begin{minipage}[b]{1\linewidth}
    \centering
    \centering{\scalebox{0.5}{\includegraphics{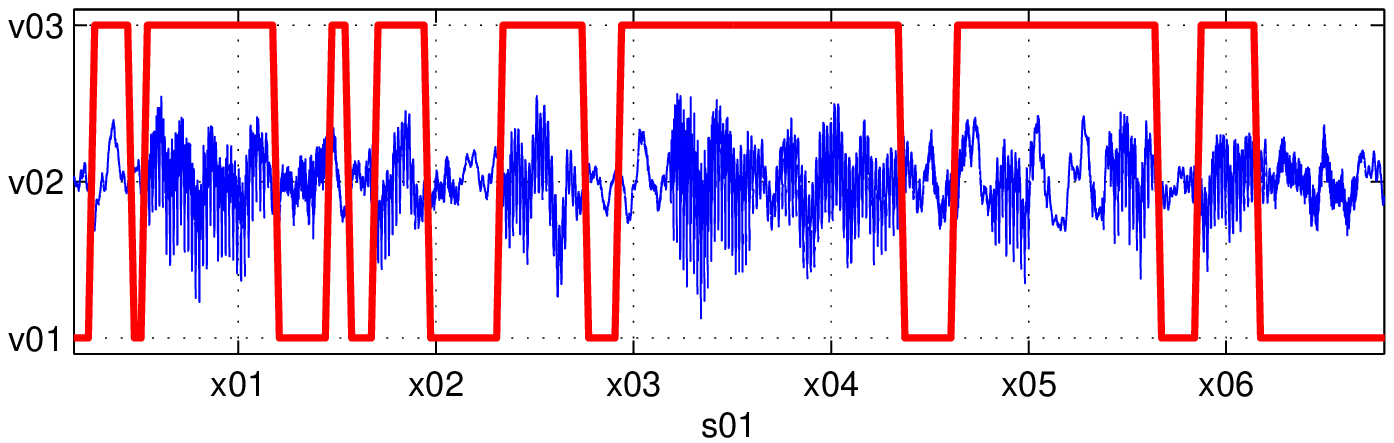}}}
    \centerline{(b) Voiced-speech classification.}
\end{minipage}
\hfill
\begin{minipage}[b]{1\linewidth}
    \centering
    \centering{\scalebox{0.5}{\includegraphics{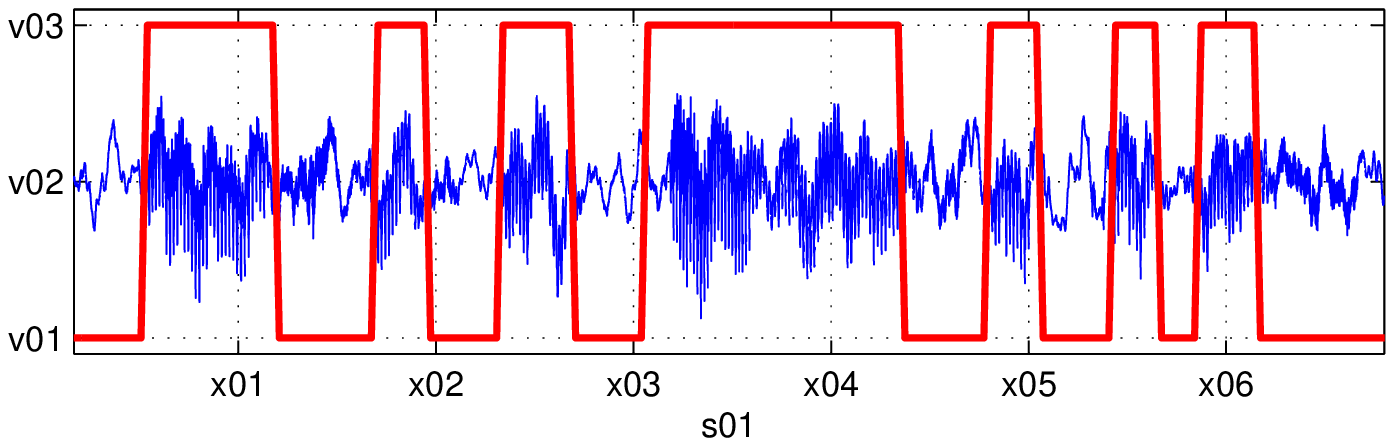}}}
    \centerline{(c) Voiced-speech classification after SpTe-ACF.}
\end{minipage}
\caption{Speech signal with additive car noise at $0$ dB SNR and classified voiced frames.}
\label{fig:vu_classifier}
\end{figure}

Fig.~\ref{fig:vu_classifier} (a) presents an example for the voiced-speech classification and the corresponding original noise-free signal waveform. 
Fig.~\ref{fig:vu_classifier} (b) shows the decision for the same signal waveform with additive car noise. Applying Eq.~(\ref{eq:decision}), 
we can observe two different types of errors: 

In some frames the speech is masked by the additive noise so that the speech frame is wrongly 
marked as unvoiced. The second error type occurs if the noise also includes periodic elements. As a result, the speech frame may be 
misclassified as voiced. We can recover some of the second type of error by post-processing, which will be explained below. 
The final voiced speech decision including post-processing is illustrated in Fig.~\ref{fig:vu_classifier} (c).

\section{Pitch Tracking}
The SpTe-ACF combines the normalized time- and frequency-domain autocorrelations to obtain an 
estimate for the current pitch value given a voiced speech frame. Subsequent to the SpTe-ACF, we apply the FB-KF to fit the entire pitch track.

\subsection{Spectro-Temporal Autocorrelation}
In post-processing, we first change the decision for those single unvoiced frames that lie between two voiced frames. 

Then, as described in~\cite{Kondoz}, we compute the time-domain autocorrelation $R_t(k,l)$ and the 
frequency-domain autocorrelation $R_s(k,l)$ for $N_{\mathrm{low}} \leq l \leq N_{\mathrm{high}}$. 
The lower and upper limits for the pitch search are chosen to be 
$ N_{\mathrm{low}} = \max \big\{ 1,\big\lfloor F_s / F_{\mathrm{max}} \big\rfloor \big\}$ and 
$N_{\mathrm{high}} = \min \big\{ \big\lceil F_s / F_{\mathrm{min}}  \big\rceil ,N \big\}$,\\ 
where $\lfloor ~ \rfloor$ and $\lceil~ \rceil$ denote downward and upward rounding, respectively, $F_s$ is the sampling frequency and 
$F_{\mathrm{min}}$, $F_{\mathrm{max}}$ define the minimum and maximum values of $F_0$. 
Now, both methods are combined as follows 
\begin{align}
    \label{STACF}
    R_{st}(k,l) =  \alpha_RR_t(k,l) + (1-\alpha_R) R_s(k,l), 
\end{align}
to obtain the SpTe-ACF. The weighting factor in Eq.~(\ref{STACF}) is chosen as $\alpha_R = 0.5$. The observation for the Kalman filter is then given by
\vspace{-0.1in}
\begin{align}
    \label{eq:inputKF}
    \qquad\qquad N(k) = \arg\max_{l} \big\{ R_{st}(k,l) \big\}.
\end{align} \vspace{-0.15in}

\noindent After applying the pitch estimation algorithm to all voiced frames, we change the voicing decision of all frames 
where the estimated pitch values lie on the boundary of the search interval, i.e. $N(k) \in\left\{ N_{\mathrm{low}},N_{\mathrm{high}} \right\}$, 
to unvoiced. Finally, we change the decision for all single voiced frames that lie between two unvoiced frames to obtain contiguous 
segments.

\subsection{Kalman Filter}
The Kalman filter is based on tracking $N_{0}(k) = \frac{F_s}{F_0(k)}$. 
Finding the correct pitch track of a speech signal is equivalent to tracking an unknown 
time-varying state variable using noisy observations. To describe the transition between consecutive states $N_0(k)$ and $N_0(k-1)$ 
we can assume a simple first-order Markov model~\cite{Haykin} to formulate the state equation $ N_0(k) = N_0(k-1) + \Delta N_0(k-1)$, 
where $\Delta N_0(k) \sim \mathscr{N}( 0, \sigma^2_{\Delta_0})$ is the system noise. 
The observation equation can be defined as $ N(k) = N_0(k) + \Delta N(k)$, 
where $\Delta N(k) \sim \mathscr{N}( 0, \sigma^2_{\Delta}(k))$  is the observation noise. The prediction step of our scalar Kalman filter 
is given by~\cite{Kay} 
\begin{align*}
    \hat{N}_0(k|k-1) &= \hat{N}_0(k-1|k-1), \\
    \hat{P}_0(k|k-1) &= \hat{P}_0(k-1|k-1) + \sigma^2_{\Delta_0}
\end{align*}  \vspace{-0.18in}

\noindent and the correction step can be written as
\vspace{-0.05in}
 \begin{align*}
    g(k) &= \frac{\hat{P}_0(k|k-1)}{\hat{P}_0(k|k-1) + \sigma^2_{\Delta}(k)}, \\
    \hat{N}_0(k|k) &= \hat{N}_0(k|k-1) + g(k) \big[ N(k) - \hat{N}_0(k|k-1) \big], \\
    \hat{P}(k|k) &= \big[ 1- g(k) \big] \hat{P}_0(k|k-1).
\end{align*}
In this context, the variance of the system noise is fixed and chosen a-priori.
We need to estimate, however, the frame-dependent uncertainty of our observation $\sigma^2_{\Delta}(k)$ in each iteration step. 
Similarly to~\cite{harmonic} the instantaneous variance is obtained by the ML estimate based on the last $L$ values of 
Eq.~(\ref{eq:inputKF}), 
\vspace{-0.15in}
\begin{align}
    \hat{\sigma}^2_{\Delta}(k) = \frac{1}{L} \sum_{i=k}^{k-L+1}{\big[  N(i) -  \hat{\mu}_N(i) \big]^2},
\end{align} \vspace*{-0.1in}

\noindent where $ \hat{\mu}_N(k)$ denotes the estimated mean value of our observation, which can be computed recursively by
\begin{align}
    \label{EqEN}
     \hat{\mu}_N(k) &= \alpha N(k) + (1-\alpha)  \hat{\mu}_N(k-1).
\end{align}
The Kalman filter is applied twice, forward and backward to obtain the respective estimated pitch values
$\hat{N}_{0,f}(k|k)$ and $\hat{N}_{0,b}(k|k)$.

\subsection{Forward-Backward Pitch Tracking}
The fusion of $\hat{N}_{0,f}(k|k)$ and $\hat{N}_{0,b}(k|k)$ follows by estimating the maximum of the log-likelihood, which is defined as
\begin{align}
    \label{eq:loglik}
    \log \mathscr{L}(N_0(k)) = \log \big\{ p(\hat{N}_{0,f}(k|k),\hat{N}_{0,b}(k|k)|N_{0}(k)) \big\} .
\end{align} \vspace{-0.15in}

\noindent For the joint distribution we assume that $\hat{N}_{0,f}$ and $\hat{N}_{0,b}$ are conditionally independent given the true state $N_{0}$,
\begin{align}
    \label{eq:joint}
    p(\hat{N}_{0,f}(k|k),\hat{N}_{0,b}(k|k)|N_{0}(k&)) 
	\nonumber \\
	= p(\hat{N}_{0,f}(k|k)|N_{0}(k))& ~p(\hat{N}_{0,b}(k|k)|N_{0}(k)).
\end{align}
If we further assume that $\hat{N}_{0,f}$ and $\hat{N}_{0,b}$ in Eq.~(\ref{eq:joint}) are Gaussian distributed,
\begin{align*}
  p(\hat{N}_{0,f}(k|k)|N_{0}(k)) &\sim \mathscr{N}( N_{0}(k), \sigma^2_{\Delta,f}(k)), \\
  p(\hat{N}_{0,b}(k|k)|N_{0}(k)) &\sim \mathscr{N}( N_{0}(k), \sigma^2_{\Delta,b}(k)),
\end{align*}
then the ML estimate of Eq.~(\ref{eq:loglik}) is given by
\begin{align}
    \label{eq:fusion}
    \hat{N}_0(k)  &= \arg \max_{N_{0}(k)} \big\{ \log \mathscr{L}(N_0(k)) \big\} \nonumber \\
		  &= \frac{ \hat{\sigma}^2_{\Delta,b}(k) \hat{N}_{0,f}(k|k)  + \hat{\sigma}^2_{\Delta,f}(k) \hat{N}_{0,b}(k|k)}
			{ \hat{\sigma}^2_{\Delta,f}(k) + \hat{\sigma}^2_{\Delta,b}(k) }.
\end{align}
Finally, the relation of the estimated fundamental frequency and Eq.~(\ref{eq:fusion}) is given by
\vspace*{-0.1in}
\begin{align}
    \label{eq:fusionF0}
    \hat{F}_0(k) = \frac{F_s}{\hat{N}_0(k)}.
\end{align}\vspace*{-0.1in}

\begin{figure}[t]
    \psfrag{s08}[c][c][1]{}
    \psfrag{s09}[c][c][1]{}
    \psfrag{s10}[c][c][1]{}
    \psfrag{s11}[c][c][1]{}
    \psfrag{x01}[c][c][1.3]{$0.3$}
    \psfrag{x02}[c][c][1.3]{$0.5$}
    \psfrag{x03}[c][c][1.3]{$0.7$}
    \psfrag{x04}[c][c][1.3]{$0.9$}
    \psfrag{x05}[c][c][1.3]{$1.1$}
  \begin{minipage}[b]{1\linewidth}
    \psfrag{s01}[t][c][1.3]{time (s)}
    \psfrag{v01}[r][r][1.3]{-$1$}
    \psfrag{v02}[r][r][1.3]{$0$}
    \psfrag{v03}[r][r][1.3]{$1$}\centering
    \centering{\scalebox{0.53}{\includegraphics{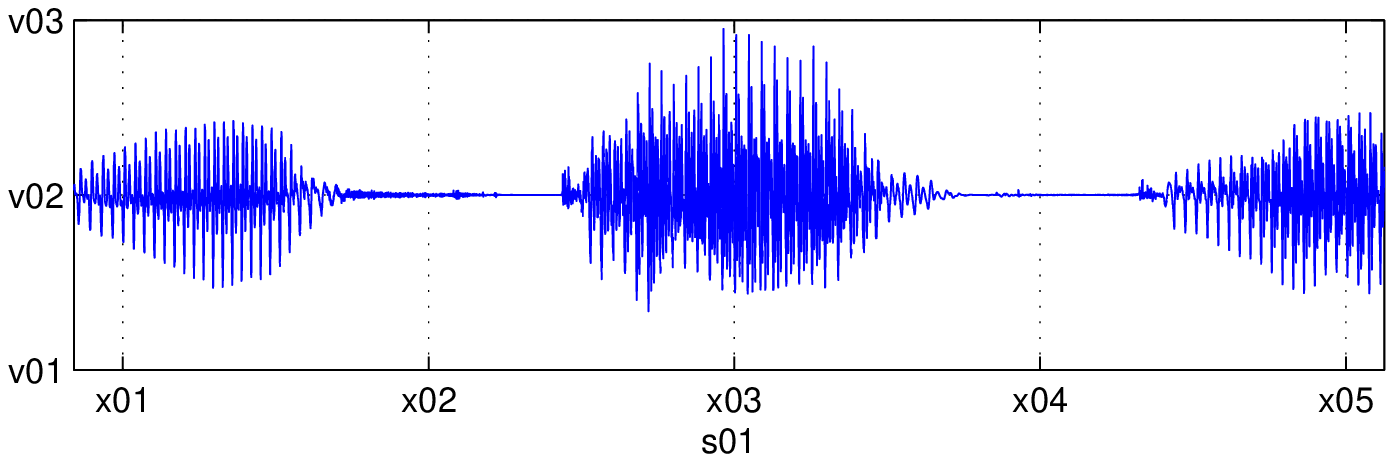}}}
    \small
    \centerline{(a) Speech waveform of male speaker.}
\end{minipage}
    \psfrag{s01}[b][c][1.3]{}
    \psfrag{s02}[t][c][1.3]{time (s)}
    \psfrag{v01}[r][r][1.3]{}
    \psfrag{v02}[r][r][1.3]{$80$}
    \psfrag{v03}[r][r][1.3]{$100$}
    \psfrag{v04}[r][r][1.3]{$120$}
    \psfrag{v05}[r][r][1.3]{$140$}
\hfill
\begin{minipage}[b]{1\linewidth}
    \centering
    \psfrag{v06}[r][r][1.3]{$160$}
    \psfrag{s06}[l][l][1.3]{$\hat{N}_{0,f}(k|k) + 10 ~\hat{\sigma}_{\Delta,f}(k)$}
    \psfrag{s07}[l][l][1.3]{$\hat{N}_{0,f}(k|k) - 10 ~\hat{\sigma}_{\Delta,f}(k)$}
    \centering{\scalebox{0.53}{\includegraphics{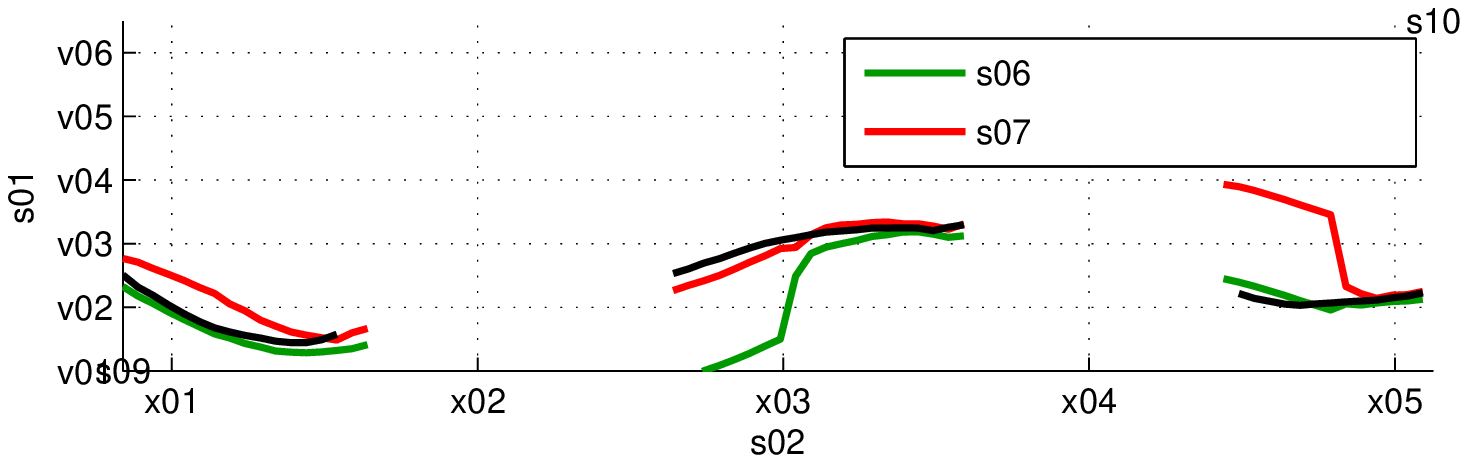}}}
    \small
    \centerline{(b) Forward Kalman filtering results with GT (black).}
\end{minipage}
\hfill
\begin{minipage}[b]{1\linewidth}
    \centering
    \psfrag{s06}[l][l][1.3]{$\hat{N}_{0,b}(k|k) + 10 ~\hat{\sigma}_{\Delta,b}(k)$}
    \psfrag{s07}[l][l][1.3]{$\hat{N}_{0,b}(k|k) - 10 ~\hat{\sigma}_{\Delta,b}(k)$}
    \centering{\scalebox{0.53}{\includegraphics{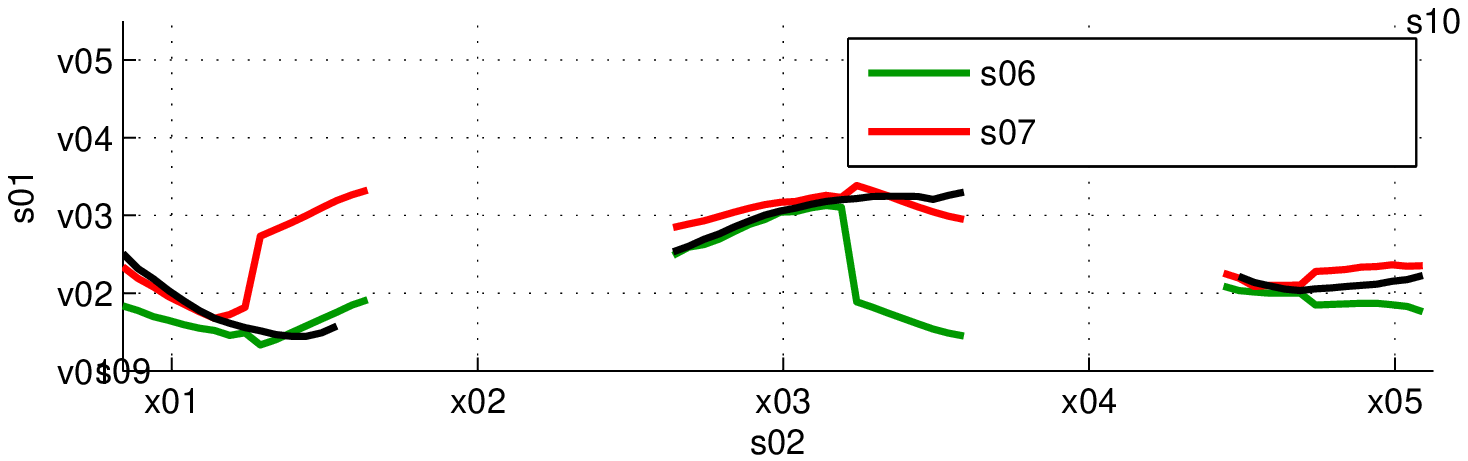}}}
    \small
    \centerline{(c) Backward Kalman filtering results with GT (black).}
\end{minipage}
\hfill
\begin{minipage}[b]{1\linewidth}
    \centering
    \psfrag{s01}[b][c][1.3]{frequency (Hz)}
    \psfrag{v01}[r][r][1.3]{$100$}
    \psfrag{v02}[r][r][1.3]{$200$}
    \psfrag{v03}[r][r][1.3]{$300$}
    \psfrag{s06}[l][l][1.3]{$\hat{F}_{0,f}(k)$}
    \psfrag{s07}[l][l][1.3]{$\hat{F}_{0,b}(k)$}
    \psfrag{s08}[l][l][1.3]{$\hat{F}_{0}(k)$}
    \centering{\scalebox{0.53}{\includegraphics{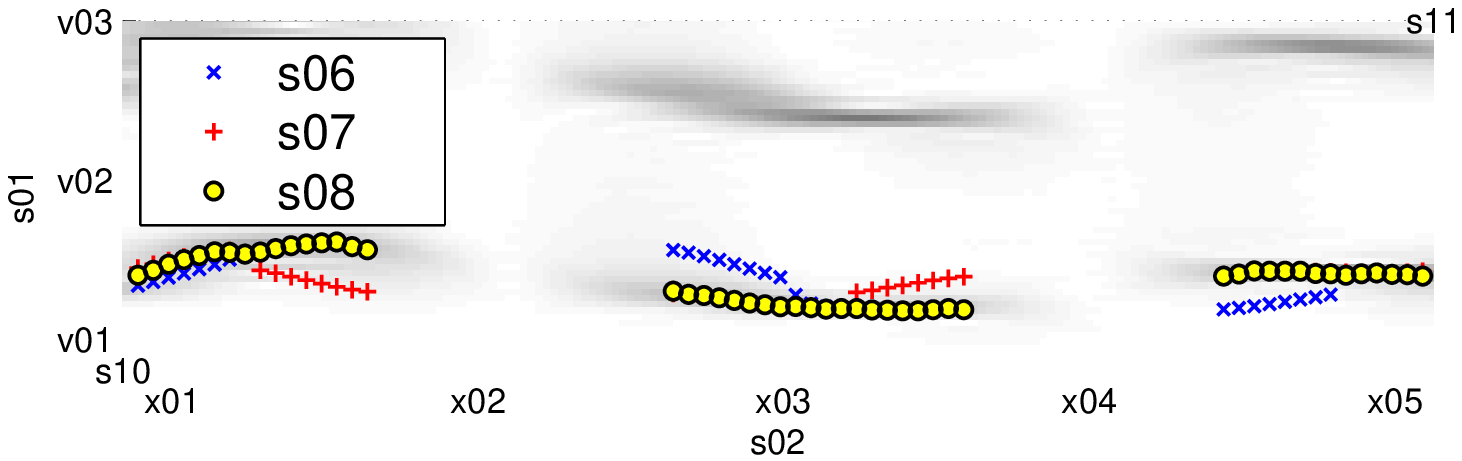}}}
    \small
    \centerline{(d) FB-Kalman pitch estimates plotted on spectrum.}
\end{minipage}
\caption{Variance-based averaging of the results of forward and backward Kalman filters.}
\label{fig:vu_kalman}
\end{figure}

Plots (b)-(d) of Fig.~\ref{fig:vu_kalman} illustrate the advantage of the FB-KF for the 
speech waveform in  Fig.~\ref{fig:vu_kalman} (a).
Starting with Fig.~\ref{fig:vu_kalman} (b), we see the curves of 
$\hat{N}_{0,f}(k|k) \pm 10 ~\hat{\sigma}_{\Delta,f}(k)$ and the groundtruth (GT), which is computed as described in 
Section~\ref{SR}.
We can observe that at the beginning or after speech pauses, the uncertainty of the observation increases. 
The increase follows from the change of $F_0$ after speech pauses.
Similarly, we observe the same results in the backward variance in Fig.~\ref{fig:vu_kalman} (c). 
Applying Eq.~(\ref{eq:fusion}) and~(\ref{eq:fusionF0}) leads to greater reliability at the beginning and end of speech segments which can be 
seen in Fig.~\ref{fig:vu_kalman} (d). See, for example, the behavior around \unit[0.7]{s}.
\begin{figure}[ht]
    \psfrag{s08}[c][c][1]{}
    \psfrag{s09}[c][c][1]{}
    \psfrag{s10}[c][c][1]{}
    \psfrag{s11}[c][c][1]{}
    \psfrag{x01}[c][c][1.3]{}
    \psfrag{x02}[c][c][1.3]{$1.5$}
    \psfrag{x03}[c][c][1.3]{$1.6$}
    \psfrag{x04}[c][c][1.3]{$1.7$}
    \psfrag{x05}[c][c][1.3]{$1.8$}
    \psfrag{x06}[c][c][1.3]{$1.9$}
    \psfrag{x07}[c][c][1.3]{$2.0$}
 \begin{minipage}[b]{1\linewidth}
    \psfrag{s01}[t][c][1.3]{time (s)}
    \psfrag{v01}[r][r][1.3]{-$0.5$}
    \psfrag{v02}[r][r][1.3]{$0$}
    \psfrag{v03}[r][r][1.3]{$0.5$}\centering
    \centering{\scalebox{0.53}{\includegraphics{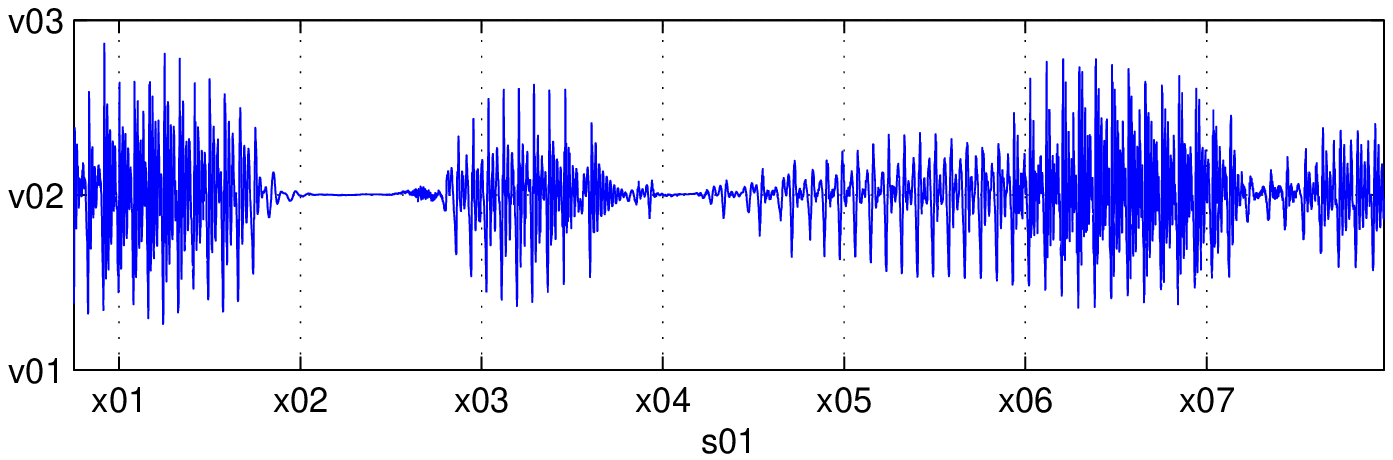}}}
    \small
    \centerline{(a) Noise-free speech waveform.}
\end{minipage}
\hfill
    \psfrag{s06}[l][l][1.3]{SpTe-ACF}
    \psfrag{s07}[l][l][1.3]{SpTe-ACF \& FB-Kalman}
    \psfrag{s09}[c][c][1]{}
    \psfrag{s10}[c][c][1]{}
    \psfrag{s01}[b][b][1.3]{frequency (Hz)}
    \psfrag{s02}[t][c][1.3]{time (s)}
    \psfrag{v01}[r][r][1.3]{$100$}
    \psfrag{v02}[r][r][1.3]{$200$}
    \psfrag{v03}[r][r][1.3]{$300$}
    \psfrag{v04}[r][r][1.3]{$400$}
\begin{minipage}[b]{1\linewidth}
    \centering{\scalebox{0.53}{\includegraphics{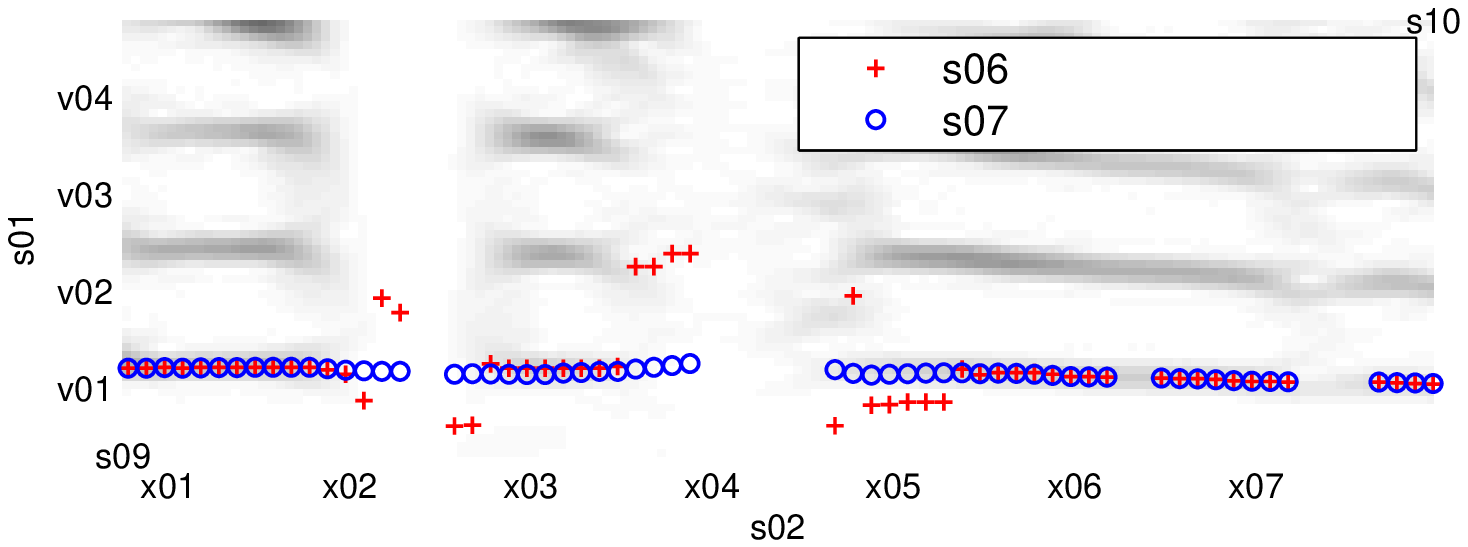}}}
    \small
    \centerline{(b) SpTe-ACF and FB-KF results plotted on noise-free spectrum.}
\end{minipage}
\caption{SpTe-ACF and FB-Kalman for $0$ dB factory noise plotted on the clean spectrum.}
\label{fig:acf_kalman}
\end{figure}

In addition, Fig.~\ref{fig:acf_kalman} shows the pitch tracking results, before and after applying the FB-KF to SpTe-ACF in a noisy environment. 
The clear speech waveform of
Fig.~\ref{fig:acf_kalman} (a) is superimposed with factory noise at $0$ dB. As a result, some pitch estimates of the SpTe-ACF differ from the 
plotted clear spectrum in the background. The FB-KF is able to correct the measurement errors here, as well. 

\vspace{-0.1in}
\section{Simulation Results}
\label{SR}
In ~\cite{BaNa,study3,Halcyon} recorded speech signals from the Keele, LDC, CSTR and PTDB-TUG databases are used to compare 
different state-of-the-art unsupervised pitch estimation algorithms. 
Here, we test the FB-KF based SpTe-ACF with the TIMIT database to find optimal parameters in terms of Gross Pitch Error (GPE) and 
(Mean Fine Pitch Error) MFPE. First we count all frames, where the relative distance between the estimated pitch value and the reference pitch value 
is higher than \unit[10]{\%},
\vspace{-0.05in}
\begin{align*}
    \mathrm{error}(k) = 
    \begin{cases} 
	\label{eq:Ind_GPE}
	      1 & \text{if} ~\frac{ |F_0(k) - \hat{F}_0(k)| }{F_0(k)} > 0.1 \\
	      0 & \text{otherwise}.
    \end{cases}
\end{align*} \vspace{-0.1in}

\noindent Then, the GPE ratio is defined as $\mathrm{GPE} = \frac{1}{K} \sum_{k=1}^K \mathrm{error}(k)$.
In the following, we only estimate the pitch values for the frames classified as voiced and calculate the corresponding GPE. 
For all voiced frames without GPE, the MFPE is defined as
\vspace{-0.1in}
\begin{align*}
    \mathrm{MFPE} = \frac{1}{N_e} \sum_{k=1}^{N_e}\frac{ |F_0(k) - \hat{F}_0(k)| }{F_0(k)},
\end{align*} \vspace{-0.1in}

\noindent where $N_e$ denotes the number of frames without GPE~\cite{Rapt}. We use Praat to determine the groundtruth $F_0(k)$.
The searching interval is defined as $F_{\mathrm{min}}=$ \unit[60]{Hz} and $F_{\mathrm{max}}=$ \unit[460]{Hz} and 
the power threshold is determined as $ \delta_{P_x} = 0.3\cdot\mathrm{E} \left\{ P_x(k) \right\}$.
The goal is to find a set of parameters $\{ L, \alpha,\sigma^2_{\Delta_0}\}$ which solves the trade-off between minimizing the 
GPE and minimizing the MFPE. The test signals show good results for both metrics if we choose $L=8$, $\alpha=0.95$, $\sigma^2_{\Delta_0}=0.06$.
If we vary the set of parameters, i.e. $L=6$, $\alpha=0.9$, $\sigma^2_{\Delta_0}=0.01$, GPE rate decreases, but this setting also 
results in an increase of MFPE rate. 

We use the ARCTIC~\cite{ARCTIC} database for evaluation, where the signals are sampled at $F_s$ = \unit[16]{kHz}. 
We compare the proposed method with Yin, Pefac, BaNa and Halcyon which are applied with their default parameters. 
First we need to calculate the GT as a reference for the GPE. A speech frame is marked to be voiced, when 
the ratio of the distance between the maximum and minimum pitch estimates with respect to the minimum value of all algorithms is smaller 
than \unit[10]{\%}. The GT value is then computed as the average of all
pitch estimates. Otherwise the frame is considered as unvoiced.
As a consequence, there are some ARCTIC signals where only single frames within the entire sentence remain. 
Therefore, we apply the unsupervised voiced speech classification method
as described in Section~\ref{sec:classification} and consider only those signals, where at least \unit[60]{\%} of voiced frames are still available.

\begin{figure}[t]
\begin{minipage}[b]{1\linewidth}
    \psfrag{s01}[t][t][1.5]{SNR (dB)}
    \psfrag{s02}[b][t][1.5]{GPE (\%)}
    \psfrag{s06}[l][l][1.3]{Yin}
    \psfrag{s07}[l][l][1.3]{Pefac}
    \psfrag{s08}[l][l][1.3]{BaNa}
    \psfrag{s09}[l][l][1.3]{Halcyon}
    \psfrag{s10}[l][l][1.3]{SpTe-ACF \& FB-Kalman}
    \psfrag{s12}[l][l][1]{}
    \psfrag{s13}[l][l][1]{}
    \psfrag{x01}[c][c][1.3]{$0$}
    \psfrag{x02}[c][c][1.3]{$2$}
    \psfrag{x03}[c][c][1.3]{$4$}
    \psfrag{x04}[c][c][1.3]{$6$}
    \psfrag{x05}[c][c][1.3]{$8$}
    \psfrag{x06}[c][c][1.3]{$10$}
    \psfrag{v05}[r][r][1.3]{$20~~$}
    \psfrag{v04}[r][r][1.3]{$15~~$}
    \psfrag{v03}[r][r][1.3]{$10~~$}
    \psfrag{v02}[r][r][1.3]{$5~~$}
    \psfrag{v01}[r][r][1.3]{$0~$}
    \centering
    \centering{\scalebox{0.53}{\includegraphics{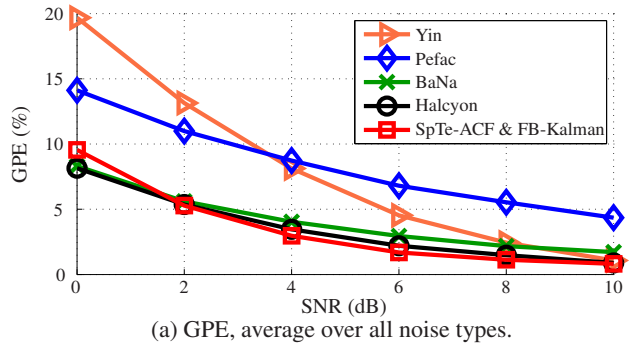}}}
    \small 
    \centerline{(a) GPE, average over all noise types.}
\end{minipage}
\hfill
\begin{minipage}[b]{1\linewidth}
    \psfrag{x01}[c][c][1.3]{$0$}
    \psfrag{x02}[c][c][1.3]{$2$}
    \psfrag{x03}[c][c][1.3]{$4$}
    \psfrag{x04}[c][c][1.3]{$6$}
    \psfrag{x05}[c][c][1.3]{$8$}
    \psfrag{x06}[c][c][1.3]{$10$}
    \psfrag{s01}[t][t][1.5]{SNR (dB)}
    \psfrag{s02}[b][t][1.5]{MFPE (\%)}
    \psfrag{s06}[l][l][1.3]{Yin}
    \psfrag{s07}[l][l][1.3]{Pefac}
    \psfrag{s08}[l][l][1.3]{BaNa}
    \psfrag{s09}[l][l][1.3]{Halcyon}
    \psfrag{s10}[l][l][1.3]{SpTe-ACF \& FB-Kalman}
    \psfrag{s12}[l][l][1]{}
    \psfrag{s13}[l][l][1]{}
    \psfrag{v07}[r][r][1.3]{$1.8~~$}
    \psfrag{v06}[r][r][1.3]{$1.6~~$}
    \psfrag{v05}[r][r][1.3]{$1.4~~$}
    \psfrag{v04}[r][r][1.3]{$1.2~~$}
    \psfrag{v03}[r][r][1.3]{$1~~$}
    \psfrag{v02}[r][r][1.3]{$0.8~~$}
    \psfrag{v01}[r][r][1.3]{$0.6~$}
    \centering
    \centering{\scalebox{0.53}{\includegraphics{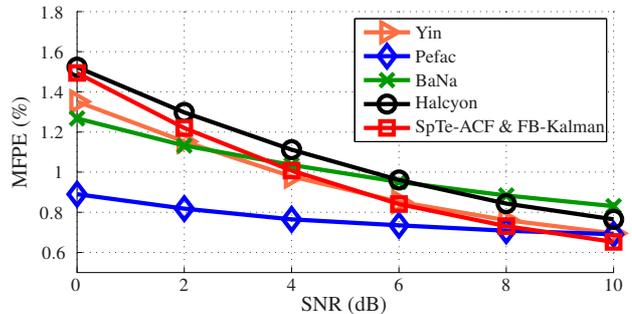}}}
    \small
    \centerline{(b) MFPE, average over all noise types.}
\end{minipage}
\label{fig:sim_res}
\vspace{-0.15in}
\caption{Simulation results for $4$ speakers, $50$ signals, $8$ noise types and $6$ different SNR values.}
\end{figure}

Fig.~\ref{fig:sim_res} presents the resulting GPE rates for different types of additive noise~\cite{NOISE} and for a range of SNR values.
Regarding the curves in Fig.~\ref{fig:sim_res} (a) for different algorithms, it can be seen, that the FB-KF based SpTe-ACF achieves 
the highest overall average detection rate for SNR values up to \unit[0]{dB}.
In addition, Fig.~\ref{fig:sim_res} (b) highlights the MFPE rates for a range of SNR values. Again, the FB-KF based SpTe-ACF achieves 
a good performance for SNR values up to \unit[0]{dB}.
In summary, we observe an acceptable trade-off between GPE and MFPE for the chosen set of parameters.

\begin{table}[ht]
  \centering
  \small
  \begin{tabular*}{0.995\columnwidth}{lccccc}
    \hline 
    &\begin{scriptsize}Yin\end{scriptsize}
    & \begin{scriptsize}Pefac\end{scriptsize}
    & \begin{scriptsize}BaNa\end{scriptsize}
    & \begin{scriptsize}Halcyon\end{scriptsize}
    & \begin{scriptsize}SpTe-ACF \& FB-KF\end{scriptsize} 
    \\
    \hline
     \begin{scriptsize}avg. time\end{scriptsize}
	  & \begin{scriptsize}\unit[0.03]{s}\end{scriptsize}
	  & \begin{scriptsize}\unit[0.64]{s}\end{scriptsize}
	  &\begin{scriptsize}\unit[4.16]{s}\end{scriptsize} 
	  &\begin{scriptsize}\unit[3.64]{s}\end{scriptsize}
	  &\begin{scriptsize}\unit[0.26]{s}\end{scriptsize}
	\\
    \hline 
  \end{tabular*}
  \caption{Average computation time in seconds for all algorithms implemented in Matlab. The average length of all 
	  speech signals is $3.07$ s. }
  \label{tab:time}
\end{table}

\vspace{-0.2in}
\section{Conclusions}
We have presented a forward-backward Kalman filtering technique for pitch tracking in noisy environments. The procedure includes a sentence-based 
method for unsupervised voiced-unvoiced classification. The combination of the results of the forward and backward KF can 
capture and correct outliers in the pitch estimate. The success of the algorithm is highly dependent on a good performance of 
the voiced-unvoiced classification stage. If the observation input for the Kalman filter contains too many outliers then the Kalman 
filter is drifting. Simulation results based on the ARCTIC database, however, show that the SpTe-ACF based FB-KF achieves a good pitch 
estimation accuracy in noisy environments for SNR values down to $0$ dB.  In addition, the computational effort for pitch estimation and tracking is 
significantly lower than that for BaNa and Halcyon. Table~\ref{tab:time} provides a very coarse comparison of computational complexity for some of the considered 
algorithms based on MATLAB execution times.


\clearpage
\small
\bibliographystyle{ieeetr}
\bibliography{refs}

\begin{thebibliography}{10}

\bibitem{VaryMartin}
P.~Vary and R.~Martin, {\em {Digital Speech Transmission: Enhancement, Coding
  And Error Concealment}}.
\newblock John Wiley \& Sons, 2006.

\bibitem{V-UV}
B.~Atal and L.~Rabiner, ``A pattern recognition approach to
  voiced-unvoiced-silence classification with applications to speech
  recognition,'' {\em IEEE Transactions on Acoustics, Speech, and Signal
  Processing}, vol.~24, pp.~201--212, Jun 1976.

\bibitem{Kondoz}
A.~M. Kondoz, {\em {Digital Speech: Coding for Low Bit Rate Communication
  Systems}}.
\newblock New York, NY, USA: John Wiley \& Sons, Inc., 2nd~ed., 2004.

\bibitem{study1}
L.~Rabiner, M.~Cheng, A.~Rosenberg, and C.~McGonegal, ``A comparative
  performance study of several pitch detection algorithms,'' {\em IEEE
  Transactions on Acoustics, Speech, and Signal Processing}, vol.~24,
  pp.~399--418, Oct 1976.

\bibitem{study2}
P.~Veprek and M.~S. Scordilis, ``Analysis, enhancement and evaluation of five
  pitch determination techniques,'' {\em Speech Communication}, vol.~37,
  no.~3-4, pp.~249--270, 2002.

\bibitem{study3}
S.~Lyudmila and I.~Yadigar, ``A {C}omparative {A}nalysis of {P}itch {D}etection
  {M}ethods {U}nder the {I}nfluence of {D}ifferent {N}oise {C}onditions,'' {\em
  Journal of Voice}, vol.~29, no.~4, pp.~410--417, 2015.

\bibitem{Praat}
P.~Boersma, ``Accurate short-term analysis of the fundamental frequency and the
  harmonics-to-noise ratio of a sampled sound,'' {\em Proceedings of the
  Institute of Phonetic Sciences}, vol.~17, no.~1193, pp.~97--110, 1993.

\bibitem{YIN}
A.~de~Cheveign{\'e} and H.~Kawahara, ``{YIN}, a fundamental frequency estimator
  for speech and music,'' {\em The Journal of the Acoustical Society of
  America}, vol.~111, no.~4, pp.~1917--1930, 2002.

\bibitem{RabinerSchafer}
L.~Rabiner and R.~Schafer, {\em {Theory and Applications of Digital Speech
  Processing}}.
\newblock Pearson, 2011.

\bibitem{HPS}
H.~Quast, O.~Schreiner, and M.~R. Schroeder, ``Robust pitch tracking in the car
  environment,'' in {\em Acoustics, Speech, and Signal Processing (ICASSP),
  2002 IEEE International Conference on}, vol.~1, pp.~I--353--I--356, May 2002.

\bibitem{Pefac}
S.~Gonzalez and M.~Brookes, ``{PEFAC} - {A} {P}itch {E}stimation {A}lgorithm
  {R}obust to {H}igh {L}evels of {N}oise,'' {\em IEEE/ACM Transactions on
  Audio, Speech, and Language Processing}, vol.~22, no.~2, pp.~518--530, 2014.

\bibitem{BaNa}
N.~Yang, H.~Ba, W.~Cai, I.~Demirkol, and W.~Heinzelman, ``Ba{N}a: {A} {N}oise
  {R}esilient {F}undamental {F}requency {D}etection {A}lgorithm for {S}peech
  and {M}usic,'' {\em IEEE/ACM Transactions on Audio, Speech, and Language
  Processing}, vol.~22, pp.~1833--1848, Dec 2014.

\bibitem{Theory}
M.~Christensen and A.~Jakobsson, {\em {Multi-Pitch Estimation}}.
\newblock Synthesis lectures on speech and audio processing, Morgan \& Claypool
  Publishers, 2009.

\bibitem{harmonic}
S.~Karimian-Azari, N.~Mohammadiha, J.~R. Jensen, and M.~G. Christensen, ``Pitch
  estimation and tracking with harmonic emphasis on the acoustic spectrum,'' in
  {\em 2015 IEEE International Conference on Acoustics, Speech and Signal
  Processing (ICASSP)}, pp.~4330--4334, 2015.

\bibitem{Halcyon}
E.~Azarov, M.~Vashkevich, and A.~Petrovsky, ``Instantaneous pitch estimation
  algorithm based on multirate sampling,'' in {\em 2016 IEEE International
  Conference on Acoustics, Speech and Signal Processing (ICASSP)},
  pp.~4970--4974, March 2016.

\bibitem{Deep}
K.~Han and D.~Wang, ``{N}eural {N}etwork {B}ased {P}itch {T}racking in {V}ery
  {N}oisy {S}peech,'' {\em IEEE/ACM Transactions on Audio, Speech, and Language
  Processing}, vol.~22, pp.~2158--2168, Dec 2014.

\bibitem{Theodoridis}
S.~Theodoridis, {\em {Machine Learning: A Bayesian and Optimization
  Perspective}}.
\newblock Academic Press, 1st~ed., 2015.

\bibitem{Haykin}
S.~Haykin, {\em {Adaptive Filter Theory}}.
\newblock Pearson, 5th~ed., 2014.

\bibitem{Kay}
S.~M. Kay, {\em {Fundamentals of Statistical Signal Processing: Estimation
  Theory}}.
\newblock Prentice Hall, Inc., 1993.

\bibitem{Rapt}
E.~Azarov, M.~Vashkevich, and A.~Petrovsky, ``Instantaneous pitch estimation
  based on {RAPT} framework,'' in {\em Signal Processing Conference (EUSIPCO),
  2012 Proceedings of the 20th European}, Aug 2012.

\bibitem{ARCTIC}
J.~Kominek and A.~W. Black, ``{CMU} {ARCTIC} databases for speech synthesis,''
  tech. rep., 2003.

\bibitem{NOISE}
A.~Varga and H.~J. Steeneken, ``{Assessment for automatic speech recognition:
  II. NOISEX-92: A database and an experiment to study the effect of additive
  noise on speech recognition systems},'' {\em Speech Communication}, vol.~12,
  no.~3, pp.~247--251, 1993.

\end{thebibliography}


\end{document}